\RequirePackage[l2tabu]{nag} 
\documentclass[11pt]{article}

\usepackage{fullpage}
\usepackage{float}
\usepackage{fancyhdr} 
\usepackage{lastpage} 
\usepackage{extramarks} 
\usepackage{courier} 
\usepackage[top=1in, bottom=1in, left=1in, right=1in]{geometry}

\usepackage[utf8]{inputenc}

\usepackage{amsfonts}
\usepackage[dvipsnames]{xcolor}
\usepackage{amsmath}

\usepackage{graphicx}

\usepackage{enumerate}

\usepackage{enumitem}

\makeatletter
\def\endthebibliography{%
	\def\@noitemerr{\@latex@warning{Empty `thebibliography' environment}}%
	\endlist
}
\makeatother

\usepackage{amsthm}

\newtheorem{thm}{Theorem}[section]

\newtheorem{pro}[thm]{Property}

\usepackage{amssymb}

\usepackage{authblk}

\makeatletter
\let\NAT@parse\undefined
\makeatother
\usepackage{hyperref}  

\title{
\vspace{-30mm} 
\small{
\normalfont This paper has been accepted for publication at the 58th Annual Conference on Decision and Control, 2019. \\ Please cite the paper as: Yue Guan, Anuradha M. Annaswamy, and H. Eric Tseng. \\ “Cumulative Prospect Theory Based Dynamic Pricing for Shared Mobility on Demand Services.” \\ \textit{2019 IEEE 58th Annual Conference on Decision and Control (CDC).} IEEE, 2019.
}\\
\vspace{10mm} 
\LARGE \bf Cumulative Prospect Theory Based Dynamic Pricing for Shared Mobility on Demand Services
}

\author[1]{Yue Guan\thanks{Corresponding author. Email: guany@mit.edu.}}
\author[1]{Anuradha M. Annaswamy}
\author[2]{H. Eric Tseng}
\affil[1]{Department of Mechanical Engineering, Massachusetts Institute of Technology}
\affil[2]{Research and Advanced Engineering, Ford Motor Company}

\begin{document}

\maketitle

\begin{abstract}
	
	Cumulative Prospect Theory (CPT) is a modeling tool widely used in behavioral economics and cognitive psychology that captures subjective decision making of individuals under risk or uncertainty. In this paper, we propose a dynamic pricing strategy for Shared Mobility on Demand Services (SMoDSs) using a passenger behavioral model based on CPT. This dynamic pricing strategy together with dynamic routing via a constrained optimization algorithm that we have developed earlier, provide a complete solution customized for SMoDS of multi-passenger transportation. The basic principles of CPT and the derivation of the passenger behavioral model in the SMoDS context are described in detail. The implications of CPT on dynamic pricing of the SMoDS are delineated using computational experiments involving passenger preferences. These implications include interpretation of the classic fourfold pattern of risk attitudes, strong risk aversion over mixed prospects, and behavioral preferences of self reference. Overall, it is argued that the use of the CPT framework corresponds to a crucial building block in designing socio-technical systems by allowing quantification of subjective decision making under risk or uncertainty that is perceived to be otherwise qualitative.
	
\end{abstract}




\paragraph{Index Terms\textnormal{:}} Cumulative Prospect Theory, Dynamic Pricing, Shared Mobility on Demand, Smart Cities, Risk Attitudes.

\section{Introduction}

Until recently, available solutions for urban transportation have been clearly binary, with the first option represented by public transportation that provides low cost and reduced flexibility and the second corresponding to private automobiles that have high cost and improved flexibility. Emergence of ride sharing platforms such as Uber, Lyft, and Didi Chuxing have changed this landscape, introducing a continuum of services at various levels of cost, flexibility, and carbon footprint. With a projection of a total number of 2 billion vehicles on roads by the year 2035 \cite{12Billio40:online}, the emergence of new concepts such as Mobility on Demand \cite{ambrosino2004demand, chong2013autonomy} are urgently needed. One such paradigm is the notion of Shared Mobility on Demand Services (SMoDSs), which consists of customized dynamic routing and dynamic pricing for multiple passengers. This paper pertains to an SMoDS that can provide a customized combination of affordability, flexibility, and carbon footprint. We build on our earlier work in \cite{guan2019dynamicrouting} and \cite{annaswamy2018transactive}, and offer a solution based on Cumulative Prospect Theory for determining dynamic tariffs.

The results of \cite{guan2019dynamicrouting} correspond to designing dynamic routes for passengers who request the SMoDS, based upon the requested pickup, drop-off locations, and a pre-specified bound on the walking distance by each passenger. An Alternating Minimization (AltMin) based algorithm was presented that optimizes a relevant time cost. The SMoDS server then offered pickup and drop-off locations as well as walking, waiting and riding times to each passenger derived via the AltMin algorithm. The notion of \textit{Transactive Control} in \cite{annaswamy2018transactive} was introduced to enable the SMoDS to offer a dynamic tariff to the passenger which can serve as an incentive for the decision on the offer. A passenger behavioral model based on Utility Theory \cite{von2007theory} was derived, with the utility of the passenger being a function of both travel times and tariff. The resulting socio-technical model that combines the passenger behavioral model and the optimization of dynamic routes was used to derive a desired probability of acceptance that led to the average estimated waiting time of passengers on the SMoDS platform being regulated around a desired value. The derivation of the actual dynamic tariffs was however not addressed and assumed to be such that the desired probability of acceptance from each passenger was realized.

The results mentioned above have two deficiencies. The first is that the passenger behavioral model is significantly more complex than that considered in \cite{annaswamy2018transactive}. Strategic decision making, adjustments based upon framing effect, loss aversion, and probability distortion are several key features related to subjective decision making of individuals when facing uncertainty, which makes classic Expected Utility Theory (EUT) inadequate. And an intrinsic feature of the SMoDS is uncertainty in the realized travel times as the route of the passenger could be updated at any time due to the need to accommodate new passengers during the current ride. An important concept that can be utilized towards a more accurate behavioral model for decision making under uncertainty is \textit{Prospect Theory} \cite{kahneman2013prospect, tversky1992advances} in general, and Cumulative Prospect Theory (CPT), in particular, where the distortion is applied to cumulative probabilities so as to avoid violations of first order stochastic dominance \cite{tversky1992advances}. The second deficiency is the lack of focus on specific dynamic tariffs related to the SMoDS. We address both of these deficiencies in this paper.  

The main contribution of this paper is a CPT based dynamic pricing strategy, where decisions of passengers are based on the subjective utility of the travel times and tariff offered by the SMoDS server. The overall framing, probability distortion, parameterization of the behavioral model, and impact of risk attributes on dynamic pricing are all discussed. Computational experiments involving passenger preferences are exploited to analyze various scenarios of passenger's risk attitudes via the proposed CPT based behavioral model.

Since being introduced by Kahneman and Tversky in 1979 \cite{kahneman2013prospect}, Prospect Theory has achieved remarkable successes in behavioral economics \cite {barberis2013thirty} and cognitive psychology \cite{arkes1985psychology}. Until recently, PT has been widely applied in engineering applications where uncertainty plays an important role, such as cloud storage defense \cite{xiao2017cloud}, energy storage of smart grids \cite{wang2014integrating}, and common-pool resource sharing \cite{hota2016fragility}. In the context of transportation, PT has been explored in \cite{han2005integrating} through a Stackelberg Games that studies the interplay between the objectives of individual travelers and that of the policy maker, and in \cite{xu2011prospect} through travelers' route choices when the travel times are uncertain and deriving the static tolls that result in the optimal system performance. Though PT has been investigated in the area of smart cities/transportations and asset pricing \cite{barberis2001prospect}, to the best of our knowledge, no prior work has been reported related to the applications of PT in SMoDS or for evaluating dynamic tariffs.

\section{Dynamic Routing and Dynamic Pricing}
\label{background}

\begin{figure}[h!]
	\centering
	\includegraphics[width=0.9\textwidth]{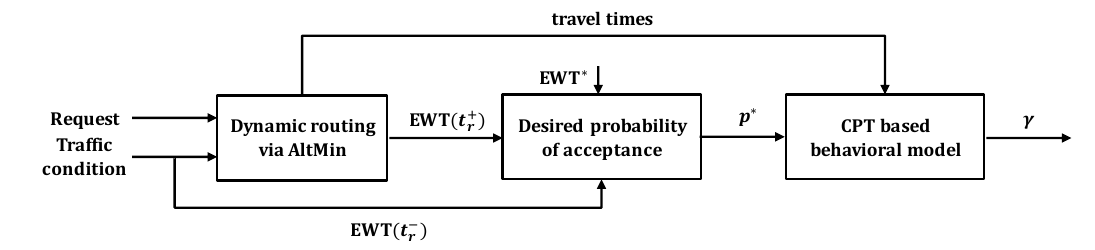}
	\caption{Overall schematic of the CPT based dynamic pricing strategy.}
	\label{fig:diagram}
\end{figure}

The problem considered in this paper is a SMoDS which accommodates ride requests from passengers in real time. The overall schematic of the CPT based dynamic pricing strategy is illustrated in Fig. \ref{fig:diagram}, which consists of three main building blocks. The first block updates the dynamic route for each passenger via the AltMin algorithm developed in \cite{guan2019dynamicrouting} when a new request is received, and calculates the updated $\text{EWT}(t)$ right after the moment of request if the passenger decides to accept the offer. $\text{EWT}(t)$ denotes the average \textit{Estimated Waiting Time} of all passengers who are in the pickup queue at timestamp $t$, i.e., have accepted the SMoDS offers but are yet to be picked up. Given the definition, it is easy to see that $\text{EWT}(t)$ can be regarded as a Key Performance Indicator (KPI) \cite{hall2015effects} to measure the degree of balance between demand and supply. We therefore apply this KPI as a desired target, ${\text{EWT}}^*$, of economic efficiency of the proposed SMoDS platform. The second block determines the desired probability of acceptance $p^*$ for the new passenger required by the SMoDS platform so as to ensure that the expected $\text{EWT}(t)$ after the passenger's decision approaches ${\text{EWT}}^*$ \cite{annaswamy2018transactive}. Finally, the third block utilizes the CPT framework to determine the dynamic tariff $\gamma$ that will nudge the passenger towards $p^*$, and forms the focus of this paper. The details of the first two blocks are described in Sections \ref{dynamicrouting} and \ref{dynamicpricing} respectively. With this overall background, we then proceed to elaborate the CPT framework starting from Section \ref{CPT}.

\subsection{Dynamic Routing via AltMin Algorithm}
\label{dynamicrouting}

An AltMin based optimization algorithm is developed in \cite{guan2019dynamicrouting} to design the optimal routes given the requested pickup, drop-off locations, and pre-specified bounds on the walking distances by the passengers, using an objective function that minimizes a weighted sum of various travel time cost terms, including the total travel time of the vehicle, the walking, waiting, and riding times of each passenger. The optimization procedure is carried out iteratively by determining a set of routing points through which the vehicle picks up and/or drops off passengers, and the sequence at which these routing points are visited. It has been demonstrated in \cite{guan2019dynamicrouting} that the AltMin algorithm is capable of accommodating real time requests, and outperforms standard Mixed Integer Quadratically Constrained Programming based approaches with an order of magnitude improvement in computational efficiency and with comparable optimality. 

\subsection{Dynamic Pricing via Utility Theory}
\label{dynamicpricing}

The behavioral model of passengers in \cite{annaswamy2018transactive} was based on utility theory and utilized to determine the probability of the passenger to accept the SMoDS offer. For this purpose, a utility function of taking any transportation option 
\begin{equation}
	\label{objective_utility}
	u = a_1 t_{\text{walk}} + a_2 t_{\text{wait}} + a_3 t_{\text{ride}} + b \gamma + c
\end{equation}
was proposed, where $t_{\text{walk}}, t_{\text{wait}}, t_{\text{ride}}$ denote the walking, waiting, riding times, respectively, $\gamma$ denotes the tariff, and $c$ denotes a constant summarizing all other unobservables that might count, such as the need for private space, the positive externalities of reducing greenhouse gas emission via sharing a trip. $a_1, a_2, a_3$ and $b$ are nonpositive weights which depend on the passenger preference regarding the transportation option. If the resulting utility is denoted as $U^{\ell}$ perceived for the SMoDS, with $U^j \in \mathbb{R}, j \in \{1, \, \cdots, \, N\}$ corresponding to the perceived utility of all $N \in {\mathbb{Z}}_{>0}$ available transportation options to choose from, the probability of accepting the SMoDS offer can be determined using discrete choice model \cite{ben1985discrete} as
\begin{equation}
	\label{discrete_choice_model}
	p^{\ell} = \frac{e^{U^{\ell}}}{\sum_{j=1}^N e^{U^j}}, \, \ell \in \{1, \, \cdots, \, N\}
\end{equation}

While (\ref{discrete_choice_model}) denotes the actual probability that the passenger will accept the SMoDS offer, from the perspective of the SMoDS platform, it is desired to provide a service that generates the desired collective performance for the platform. Let $\text{EWT}(t^-)$ and $\text{EWT}(t^+)$ denote the value of $\text{EWT}(t)$ immediately before timestamp $t$, and right after timestamp $t$ if the new passenger takes the offer. With this in mind, for the request received at $t_r$, a desired probability of acceptance $p^*$ was chosen to be a function of $\Delta \text{EWT}({t_r}^+) = \text{EWT}({t_r}^+) - {\text{EWT}}^*$ such that $\Delta \text{EWT}(t)$ was regulated around zero after $t_r$ in \cite{annaswamy2018transactive}. In general one can design $p^*$ as 
\begin{equation}
	\label{mapping}
	p^* = H\big[\text{EWT}({t_r}^-), \text{EWT}({t_r}^+) \, \big | \, {\text{EWT}}^*\big]
\end{equation} 
with the mapping $H(\cdot, \cdot | \cdot)$ designed such that the expected $\text{EWT}({t_r}^+)$ after the decision of the passenger approaches ${\text{EWT}}^*$. It was demonstrated in \cite{annaswamy2018transactive} that $H(\cdot, \cdot | \cdot)$ can be chosen such that an overall acceptance rate close to 80\% can be realized along with managing to regulate $\text{EWT}(t)$ around ${\text{EWT}}^*$. This is comparable to the statistics of 60-70\% reported in other ride sharing platforms \cite{cohen2016using}. We will therefore attempt to design the dynamic tariffs for the SMoDS to have the passenger's actual probability of acceptance defined in (\ref{discrete_choice_model}) towards the targeted value $p^*$.

\section{Behavioral Model using CPT}
\label{CPT}

\begin{figure}[h!]
	\centering
	\includegraphics[width=0.8\textwidth]{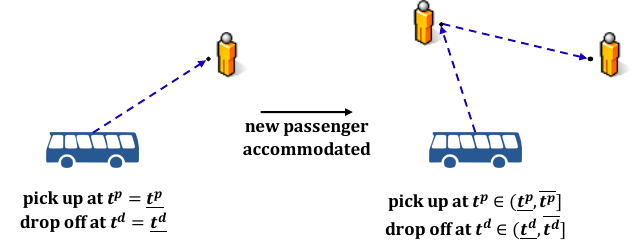}
	\caption{Source of uncertainty in the SMoDS. $t^p$ and $t^d$ denote the actual pickup and drop-off time respectively, $\underline{t^p} < \overline{t^p}$ and $\underline{t^d} < \overline{t^d}$ denote the possibly earliest and latest timestamps.}
	\label{fig:uncertainty}
\end{figure}

An important feature in the SMoDS is the presence of uncertainty, as the vehicle has to accommodate new passengers at anytime in the route. As a result, the scheduled pickup and drop-off times for a given passenger may stochastically vary over an interval (see Fig. \ref{fig:uncertainty}), making the SMoDS an uncertain prospect, i.e., a prospect with stochastic outcome, which leads to the usage of CPT. In contrast, certain prospects are ones whose outcomes are always deterministic.

The key axioms of CPT state that when making decision under uncertainty, individuals normally perceive the utility in a subjective and irrational fashion influenced by the following: \cite{kahneman2013prospect, tversky1992advances}  
\begin{itemize}
	\item \textit{Framing effect}: Individuals value prospects with respect to a reference point instead of an absolute value, and perceive gains and losses differently. 
	\item \textit{Loss aversion}: Individuals are affected much more by losses than gains. 
	\item \textit{Diminishing sensitivity}: In both gain and loss regimes, sensitivity diminishes when the prospect gets farther from the reference. Therefore, the perceived value is concave in the gain regime and convex for losses.    
	\item \textit{Probability distortion}: Individuals overweight small probability events and underweight large probability events.
\end{itemize}

\begin{figure}[h!]
	\centering
	\includegraphics[width=0.9\textwidth]{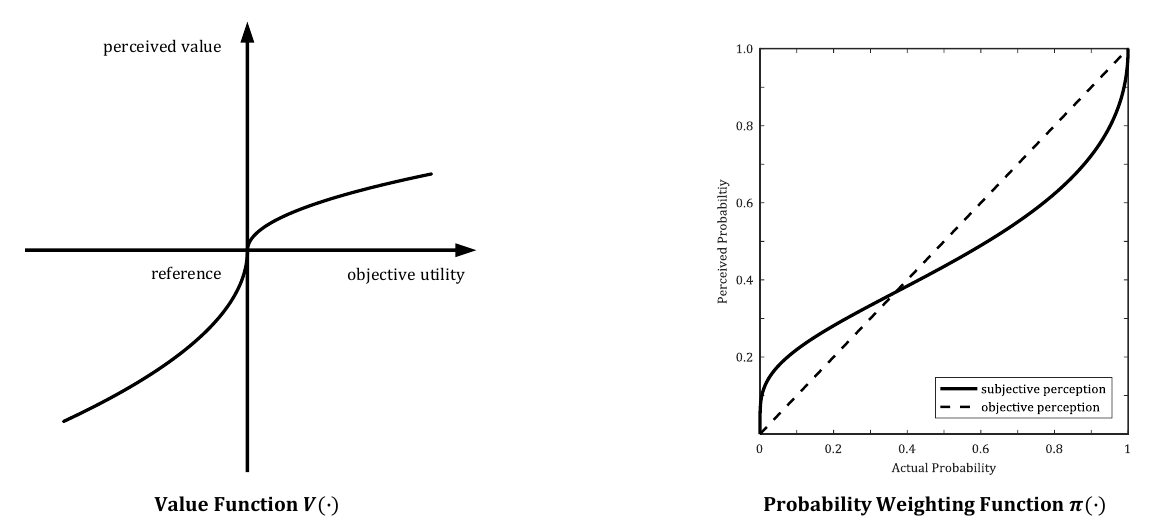}
	\caption{Illustrations of $V(\cdot)$ and $\pi(\cdot)$ in the CPT framework.}
	\label{fig:CPT_figures}
\end{figure}

A quantitative description of theses axioms is enabled by defining $V(\cdot)$ the value function and $\pi(\cdot)$ the probability weighting function, both of which are illustrated in Fig. \ref{fig:CPT_figures}. The details of the two functions are elaborated as follows. 

We first define $U$ as a random variable to denote the objective utility of an uncertain prospect, and $F_U(u)$ as the corresponding Cumulative Distribution Function (CDF). If $U$ takes on discrete values $u_i \in \mathbb{R}, \forall i \in \{1, \, \, \dots, \, n\}$ and $u_1 < \cdots < u_n$, where $n \in {\mathbb{Z}}_{>0}$ is the number of possible outcomes, one can determine the objective utility $U^o$ as the expectation of $U$ according to EUT \cite{von2007theory}, i.e.,
\begin{equation}
	\label{EUT_calculate_discrete}
	U^o = \sum_{i=1}^n p_i u_i 
\end{equation} 
where $p_i \in (0, 1)$ is the probability of outcome $u_i$, and $\sum_{i=1}^n p_i=1$. The subjective utility $U^s_R$ perceived by the passenger within the CPT framework is given by
\begin{equation}
	\label{CPT_calculate_discrete}
	U^s_R = \sum_{i=1}^n w_i V(u_i)
\end{equation}  
where $R$ denotes the reference corresponding to the framing effect\footnote{The parametrization of $R$ is marked explicitly in the subscript as the remaining discussions are heavily related to the impact of reference points.}, and $w_i$ denotes the weighting that represents the subjective perception of $p_i$. Suppose that $k$ out of the $n$ outcomes are losses, $0 \leq k \leq n, k \in \mathbb{Z}_{\geq 0}$, and the rest are non-losses, i.e., $u_i < R$ if $1 \leq i \leq k$ and $u_i \geq R$ if $k < i \leq n$, then
\begin{equation}
	\label{weights}
	w_i =
		\begin{cases}
			\pi\big[F_U(u_i)\big] - \pi\big[F_U(u_{i-1})\big], & \text{if} \quad i \in [1, k] \\
			\pi\big[1-F_U(u_{i-1})\big] - \pi\big[1-F_U(u_i)\big], & \text{otherwise}
		\end{cases} 
\end{equation} 
where we let $F_U(u_0) = 0$ for ease of notation.

In what follows, we will adopt the representations for $V(\cdot)$ and $\pi(\cdot)$ as in \cite{tversky1992advances} and \cite{prelec1998probability}, given by 
\begin{equation}
	\label{V_function}
	V(u) = 
		\begin{cases}
		{(u-R)}^{{\beta}^+}, & \text{if} \quad u \geq R \\
		-\lambda{(R-u)}^{{\beta}^-}, & \text{otherwise}
		\end{cases} 
\end{equation}
\begin{equation}
	\label{pi_function}
	\pi(p) = e^{-{[-\text{ln}(p)]}^{\alpha}}
\end{equation}
It is clear that in contrast to $U^o$, $U^s_R$ is centered on $R$, loss aversion is captured by choosing $\lambda > 1$, and diminishing sensitivity by choosing $0< {\beta}^+, {\beta}^- < 1$. The probability distortion is quantified by choosing $0 < \alpha < 1$. The extension from (\ref{CPT_calculate_discrete}) to the continuous case of $U^s_R$ is 
\begin{equation}
	\label{CPT_calculate}
	U^s_R = \int_{-\infty}^{R} V(u) \frac{d}{du}\Big\{\pi\big[F_U(u)\big]\Big\}du + \int_{R}^{\infty} V(u)\frac{d}{du}\Big\{-\pi\big[1-F_U(u)\big]\Big\}du
\end{equation}

\section{CPT based Passenger Behavioral Model in SMoDS}
\label{formulation}

The overall passenger behavioral model that we will derive in this section consists of a subjectively perceived utility $U^s_R$ and a subjective probability of acceptance $p^s_R$, both of which will be determined using CPT. The interpretation of risk attitudes, reference points, subjective weighting of probability distributions, and key properties of CPT in the SMoDS context are the topics of Sections \ref{preliminaries} through \ref{properties}. 

\subsection{Objective and Subjective Utilities} 
\label{preliminaries}

The starting point of deriving $U^s_R$ for the SMoDS is the determination of possible outcomes of its objective utility. In order to accommodate the stochastic aspects of travel times, the possible realization of the objective utility $u$ in (\ref{objective_utility}) is replaced by a random variable 
\begin{equation}
	\label{define_X}
	U = X + b \gamma
\end{equation}
where $b \gamma$ depends on the tariff from the SMoDS ride offer and is deterministic once the offer is given, and 
\begin{equation} 
	\label{calculate_X}
	X=a_1T_{\text{walk}}+a_2T_{\text{wait}}+a_3T_{\text{ride}}+c
\end{equation}
captures the uncertainty in travel times and is stochastic. Each term of the travel times is assumed to lie within a known interval specified by the SMoDS offer, defined as $T_{\text{walk}} \in [\underline{t_{\text{walk}}}, \overline{t_{\text{walk}}}], T_{\text{wait}} \in [\underline{t_{\text{wait}}}, \overline{t_{\text{wait}}}], T_{\text{ride}} \in [\underline{t_{\text{ride}}}, \overline{t_{\text{ride}}}]$. From these bounds one can determine $\underline{x}$ and $\overline{x}$, which correspond to the worst and the best cases of the travel times, respectively, and $X \in [\underline{x}, \overline{x}]$ with the CDF $F_X(x)$. Note that $F_X(x) = F_U(x+b\gamma)$ from (\ref{define_X}). With $U$ defined in (\ref{define_X})-(\ref{calculate_X}), the subjective utility $U^s_R$ is calculated via (\ref{CPT_calculate_discrete})-(\ref{calculate_X}), and objective utility $U^o$ as in (\ref{EUT_calculate_discrete}). The dependence of $U^s_R$ on $R$ is described in Section \ref{references}.  

\subsection{Interpretation of Risk Attitudes} 
\label{attitudes}

As has been shown in (\ref{discrete_choice_model}), the evaluation of the probability of acceptance requires the utility of the alternative transportation options available to the passenger. Without loss of generality, each passenger is assumed to choose between two options, the SMoDS and another option such as public transportation, UberX, which is considered as a certain prospect\footnote{The source of uncertainty such as unexpected traffic jams are small compared with that of the SMoDS, hence assumed to be negligible.} therefore with objective utility being a constant $A^o \in \mathbb{R}$. The objective probability of acceptance is given by 
\begin{equation}
	\label{prob_a_binary_objective}
	p ^ o = \frac{e ^ {U ^ o}}{e^{U^o}+e^{A^o}}
\end{equation} 
where $A^o$ can be calculated using (\ref{objective_utility}). The subjective probability of acceptance is given by 
\begin{equation}
	\label{prob_a_binary_subjective}
	p^s_R=\frac{e^{U^s_R}}{e^{U^s_R}+e^{A^s_R}}
\end{equation}  
where $A^s_R$ denotes the subjective utility of the alternative perceived by the passenger, which can be derived via (\ref{objective_utility}), (\ref{CPT_calculate_discrete}), and (\ref{V_function}). 

We now interpret the risk attitudes of passengers based on the above objective and subjective probabilities of acceptance. Since the alternative is certain, a higher probability of acceptance indicates an attitude that is more risk seeking. A passenger who is inclined to choose $p^o$ is regarded as rational. If $p^s_R > p^o$, a passenger is said to be risk seeking compared with rational passengers, and for any two references $R_1$ and $R_2$, if $p^s_{R_1} < p^s_{R_2}$, the passenger with reference $R_1$ is said to be more risk averse than the passenger with reference $R_2$, and risk seeking if the inequality is reversed.   

\subsection{Reference Points}
\label{references}

The central parameter related to CPT is $R$, and is discussed in this subsection. Three different categories are considered:
\begin{enumerate}[label = (\roman*)]
	\item \textit{Static reference points}: These correspond to any fixed quantities that are independent on the SMoDS offer. Examples include the objective utility of the alternative, i.e., $R = A ^ o$, or the utility of making the trip itself, independent of the transportation modes, to the passengers. 
	\item \textit{Dynamic reference points}: Here $R$ is dependent on the uncertain prospect itself. In the SMoDS context, $R$ can be chosen as $R = \tilde{x} + b\gamma$, where $\tilde{x}$ could be $\underline{x}, \overline{x}, \mathbb{E}_{f_X}(X)$, or any statistics preferred by the passenger. All of these examples however still correspond to deterministic references. 
	\item \textit{Stochastic reference points}: Instead of the above two categories, it is possible for the reference point itself to vary stochastically. However, little evidence has been found that supports the usage of this case \cite{baillon2017searching}, and hence we do not consider it in the rest of the paper. 
\end{enumerate}

\subsection{Subjective Weighting of Probability Distributions}
\label{distribution}

In this subsection, we discuss the subjective perception of a probability distribution $f_X(x)$ by the passenger. $f_X(x)$ denotes Probability Mass Function (PMF) if $X$ is discrete, or Probability Density Function (PDF) if $X$ is continuous. In the current problem, $f_X(x)$ represents the passenger's prediction on how long the actual travel times will be within the given intervals offered by the SMoDS server. Therefore $f_X(x)$ is objective and based upon the passenger's prior experience and assessments of demand at the time of request. In what follows, we address the subjective perception of $f_X(x)$ in both discrete and continuous cases.

\subsubsection{Continuous Distributions}

In some cases, the underlying distribution can be a truncated Normal distribution of the form
\begin{equation}
	\label{define_normal}
	f_X^n(x) = \frac{1}{Z^n}\frac{1}{\sqrt{2\pi{\sigma}^2}} e ^ {- \frac{{(x-\mu)}^2}{2{\sigma}^2}}, \, x \in [\underline{x}, \overline{x}]
\end{equation}
where $\mu=\frac{\underline{x} + \overline{x}}{2}$ and $\sigma = \overline{x} - \underline{x}$ denote the mean and standard deviation, respectively, and $Z^n = \int_{\underline{x}}^{\overline{x}} \frac{1}{\sqrt{2\pi{\sigma}^2}} e ^ {- \frac{{(x-\mu)}^2}{2{\sigma}^2}} dx > 0$ is defined for normalization.

In some other cases, a truncated exponential distribution may be valid. These are given by
\begin{equation}
	\label{define_exponential_optimistic}
	f_X^{e, o}(x) = \frac{1}{Z^{e, o}} {\lambda}^o e^{-{\lambda}^o(\overline{x}-x)}, \, x \in [\underline{x}, \overline{x}]
\end{equation}
\begin{equation}
	\label{define_exponential_pessimistic}
	f_X^{e, p}(x) = \frac{1}{Z^{e, p}} {\lambda}^p e^{-{\lambda}^p(x-\underline{x})}, \, x \in [\underline{x}, \overline{x}]
\end{equation}
where ${\lambda}^o = {\lambda}^p = \frac{1}{\overline{x} - \underline{x}}$, and $Z^{e, o} = \int_{\underline{x}}^{\overline{x}} {\lambda}^o e^{-{\lambda}^o(\overline{x}-x)} dx, Z^{e, p} = \int_{\underline{x}}^{\overline{x}} {\lambda}^p e^{-{\lambda}^p(x-\underline{x})} > 0$ are normalization constants. (\ref{define_exponential_optimistic}) and (\ref{define_exponential_pessimistic}) correspond to an optimistic and a pessimistic subcase, since the corresponding mode is at $\overline{x}$ and $\underline{x}$, respectively.

\subsubsection{Discrete Distributions}

A reasonable choice for this case is a truncated Poisson distribution of the form 
\begin{equation}
	\label{define_Poisson}
	f_X^P(x)=
		\begin{cases}
			\frac{1}{Z^P}\frac{{{(\lambda}^P)}^k e^{-{\lambda}^P}}{k!}, \, & \text{if} \, x = \overline{x} - k \frac{\overline{x}-\underline{x}}{K}\\
			0, & \text{otherwise}
		\end{cases}
\end{equation}
where $K \in {\mathbb{Z}_{>0}}$ and $k \in \{0, \, \dots, \, K\}$ denote the maximum and the actual number of possible delays, respectively, ${\lambda}^P>0$, and $Z^P = \sum_{k=0}^{K} \frac{{{(\lambda}^P)}^k e^{-{\lambda}^P}}{k!} > 0$ is the normalization constant. The truncated Poisson distribution reflects the number of possible delays due to accommodating new passengers during the ride. Each additional delay is assumed to result in the same marginal increase in travel times, hence the support of $f_X^P(x)$ is $(K+1)$ disjoint points uniformly spaced in $[\underline{x}, \overline{x}]$. The values of $K$ and ${\lambda}^P$ will be specified in Section \ref{experiments}.  

With the objective probability distributions $f_X(x)$ defined in (\ref{define_normal})-(\ref{define_Poisson}), and the reference $R$ specified, the subjective probability weighting can be derived using (\ref{weights}), and (\ref{pi_function})-(\ref{CPT_calculate}). In turn, $U^s_R$ and $p^s_R$ can be derived using (\ref{CPT_calculate_discrete}), (\ref{CPT_calculate}) and (\ref{prob_a_binary_subjective}), which completely specify the behavioral model of an SMoDS passenger.

\subsection{Key Properties of CPT based Behavioral Model}
\label{properties}

With the subjective utilities, risk attitudes, reference points, and subjective weighting of probability distributions delineated as above, we now derive four properties of the overall passenger behavioral model.

The first two properties are related to static and dynamic references, and are stated in Property \ref{monotonic_static} and \ref{monotonic_dynamic} below. These are helpful in determining the dynamic tariff $\gamma$ that allows $p^s_R$ to reach $p^*$, the desired probability of acceptance. 
\begin{pro}
	\label{monotonic_static}
	Given any static reference point $R \in \mathbb{R}$, $p^s_R$ strictly decreases with $\gamma$.
\end{pro}
\begin{pro}
	\label{monotonic_dynamic}
	Given any dynamic reference point in the form of $R = \tilde{x}+b\gamma, \tilde{x} \in \mathbb{R}$, $p^s_R$ strictly decreases with $\gamma$.
\end{pro}

Let $\bar{U} = \mathbb{E}_{f_U}(U)$ and $\bar{X} = {\mathbb{E}}_{f_X}(X)$, the third and fourth property stated in Property \ref{existence_lambda} and \ref{prob_a_for_mixed} are related to $U^s_{\bar{U}}$ and $p^s_{\bar{U}}$, respectively.
\begin{pro}
	\label{existence_lambda}
	Given any uncertain prospect, there exists a ${\lambda}^*$, such that $\forall \lambda > {\lambda}^*$, $U^s_{\bar{U}} < 0$.
\end{pro}

\begin{pro}
	\label{prob_a_for_mixed}
	For any uncertain prospect, given that $\lambda$ is sufficiently large such that $U^s_{\bar{U}} < 0$, within the price range $\gamma \in [\underline{\gamma}, \overline{\gamma})$, where $\underline{\gamma}$ satisfies $\bar{X} + b \underline{\gamma} = A^o$, and $\overline{\gamma}$ satisfies ${\big[A^o - (\bar{X} + b \overline{\gamma})\big]} ^ {{\beta}^+} - U^s_{\bar{U}} = A^o - (\bar{X} + b \overline{\gamma})$, $p^s_{\bar{U}} < p^o$.
\end{pro}

\section{Implications of CPT using Computational Experiments}
\label{experiments}

In this section, three different implications are drawn using computational experiments in order to illustrate subjective decision making of passengers, and how they can be utilized to develop the dynamic pricing strategy for the SMoDS.

\subsection{Determination of Parameters}
\label{parameterization}

The discussions in Sections \ref{background} through \ref{formulation} show that a number of parameters related to the CPT framework have to be determined. These include $\alpha, {\beta}^+, {\beta}^-, \lambda$ defined in $V(\cdot)$ and $\pi(\cdot)$, $a_1, a_2, a_3, b, c$ utility coefficients defined in (\ref{objective_utility}), $t_{\text{walk}}, t_{\text{wait}}, t_{\text{ride}}$ travel times of both the SMoDS and the alternative, and the tariff of the alternative. 

\begin{figure}[h!]
	\centering
	\includegraphics[width=0.9\textwidth]{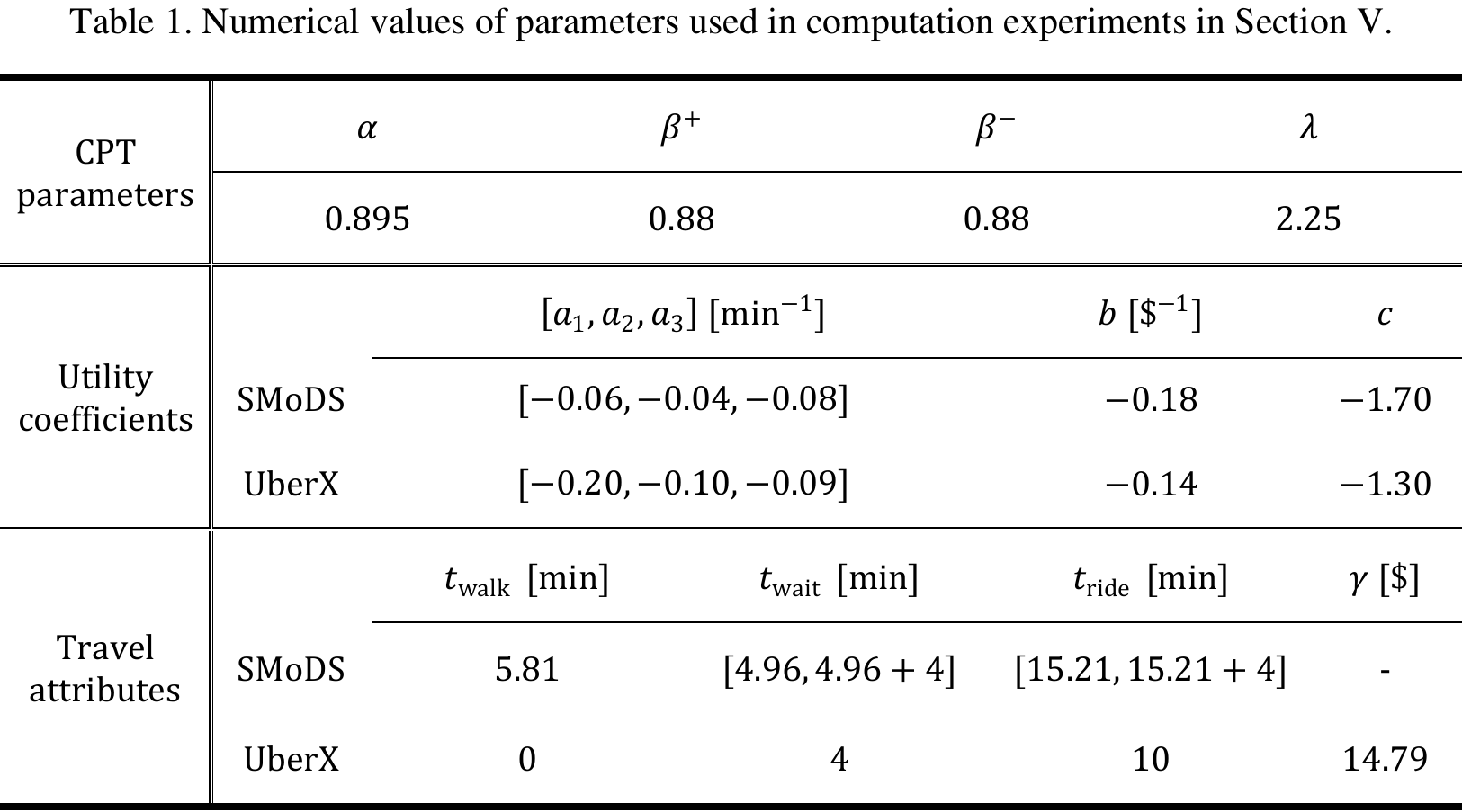}
\end{figure}
Table 1 summarizes the values of the parameters that we used in order to carry out the studies reported in this section. In particular, $\alpha$ was estimated from a recent survey study on passenger preferences under risk regarding transportation options conducted in Singapore involving $1,142$ participants with various demographics \cite{wang2018risk}, and $\beta^+, \beta^-, \lambda$ are from \cite{tversky1992advances}. In what follows, UberX is regarded as the alternative. The utility coefficients $[a_1, a_2, a_3, b, c]$ of both SMoDS and UberX were estimated from the same survey study in \cite{wang2018risk}. A dynamic routing problem of sixteen passengers using real request data from San Francisco was considered (see \cite{annaswamy2018transactive} for details), and the request from the $6^{\text{th}}$ passenger was used for the computational experiments in this section. The AltMin algorithm developed in \cite{guan2019dynamicrouting} was applied to derive the route and therefore the corresponding travel times of the SMoDS. The constraints on the possible delay were set to be at most 4 minutes of extra waiting and riding, respectively. For the same request, the travel times and price of UberX was retrieved from \cite{UberEarn32:online}. 

Using the utility coefficients, travel times and price listed in Table 1, the objective utility of UberX $A^o = -5.17$, and $\underline{x} = -3.47$, $\overline{x}=-3.07$ of the SMoDS are calculated, using (\ref{objective_utility}) and (\ref{calculate_X}). Note that $A^o, \underline{x}, \overline{x}$ are negative as they represent travel costs. 

With the above numerical values in place, we explore the three implications: (i) fourfold pattern of risk attitudes, (ii) strong aversion of mixed prospects, and (iii) self reference.

\subsection{Fourfold Pattern of Risk Attitudes}
\label{fourfoldpattern}

The fourfold pattern of risk attitudes is regarded as “the most distinctive implication of prospect theory” by Tversky and Kahneman \cite{tversky1992advances}, which states that when facing an uncertain prospect, the risk attitudes of individuals can be grouped into four categories: 
\begin{enumerate}[label = (\alph*)]
	\item Risk averse over high probability gains.
	\item Risk seeking over high probability losses.
	\item Risk seeking over low probability gains.
	\item Risk averse over low probability losses. 
\end{enumerate}
These risk attitudes are often used to justify subjective decision making of individuals for problems such as settlements of civil lawsuits, desperate treatments of terminal illnesses, playing lotteries, and getting insurance coverage. 

\begin{figure}[h!]
	\centering
	\includegraphics[width=0.9\textwidth]{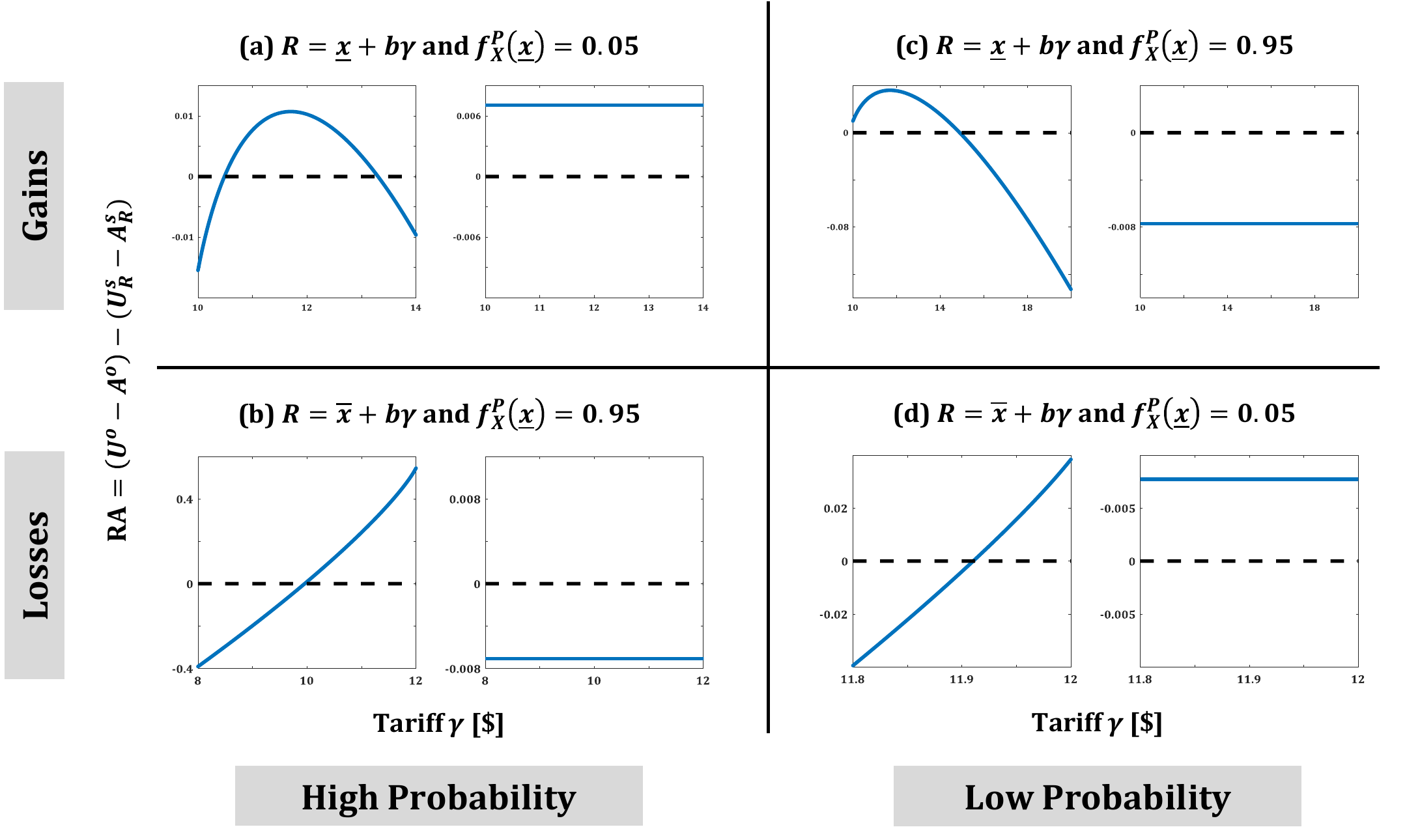}
	\caption{Illustration of the fourfold pattern of risk attitudes in the SMoDS context.}
	\label{fig:fourfold}
\end{figure}

We now illustrate the fourfold pattern in the SMoDS context using the following scenario, which corresponds to the classic setup for the analysis of the fourfold pattern \cite{tversky1992advances}: Individuals decide between two options, a certain prospect and an uncertain prospect with two outcomes. The uncertain prospect is the SMoDS, which we assume obeys a truncated Poisson distribution with $K = 1$, i.e., the passenger is subject to at most one delay. Therefore, the two possible outcomes of the SMoDS are $(\underline{x} + b\gamma) $ and $(\overline{x} + b \gamma)$. The corresponding probabilities can be determined using (\ref{define_Poisson}) as 
\begin{equation}
	\label{fourfold}
	f_X^P(\underline{x}) = \frac{{\lambda}^P}{{\lambda}^P+1}, \quad f_X^P(\overline{x}) = \frac{1}{{\lambda}^P+1}
\end{equation}
The four scenarios above are realized through suitable choices of $R$ and ${\lambda}^P$ as follows. A dynamic reference point $R$ is chosen to be either $(\underline{x} + b\gamma)$ or $(\overline{x} + b\gamma)$, the SMoDS is a gain if $R = \underline{x} + b\gamma $ and a loss if $R = \overline{x} + b \gamma$. The SMoDS is considered high probability or low probability when the outcome that is not regarded as the reference can be realized with a probability of $p_{\text{NR}}$ or $(1-p_{\text{NR}})$ respectively, where $p_{\text{NR}}$ is close to 1. In the computational experiments presented in Fig. \ref{fig:fourfold}, $p_{\text{NR}}=0.95$. Moreover, the range of the tariff is chosen as follows 
\begin{equation}
	\label{price_in_fourfold}
	\begin{cases}
		\underline{x} + b\gamma < A^o & \text{if} \; R = \underline{x} + b\gamma \\
		\overline{x} + b\gamma > A^o & \text{if} \; R = \overline{x} + b\gamma
	\end{cases}	
\end{equation}
such that the objective utility of the certain prospect, $A^o$, lies in the same gain or loss regime as the SMoDS and therefore represents a reasonable alternative to the SMoDS.

With the uncertain and the certain prospect defined in the SMoDS context above, we illustrate the fourfold pattern in Fig. \ref{fig:fourfold} using four quadrants. According to the fourfold pattern (a)-(d), the diagonal quadrants should correspond to risk averse behavior while the off-diagonal ones are risk seeking. In each quadrant, we plot a metric defined as $\text{RA}=(U^o - A^o) - (U^s_R- A^s_R)$ with respect to the tariff $\gamma$. This metric captures the Relative Attractiveness that the uncertain prospect has over the certain prospect for rational individuals versus individuals modeled with CPT. This follows since according to (\ref{prob_a_binary_objective}) and (\ref{prob_a_binary_subjective}), $\text{RA} >0 \Rightarrow p^o > p^s_R$. In Fig. \ref{fig:fourfold}, we note that $\text{RA}>0$ corresponds to all regions where the blue curve is above zero and indicates risk averse attitudes, as rational individuals have higher probability to accept the uncertain prospect than irrational ones. Similarly, $\text{RA} < 0$ corresponds to the blue line being below zero and denotes risk seeking attitudes. In each quadrant, two subplots are provided, where the subplot on the right corresponds to a specific set of parameters $\beta^+ = \beta^- = \lambda = 1$ which completely removes the role of $V(\cdot)$, while the subplot on the left corresponds to all CPT parameters chosen as in Table 1, and therefore a general CPT model. And as explained before, each quadrant corresponds to a specific choice of $R$ and ${\lambda}^P$, which together determine if an outcome is a gain or loss, and is high or low probability. 

The most important observation from Fig. \ref{fig:fourfold} comes from the differences between the left and right subplots in each of the four quadrants. For example, from Fig. \ref{fig:fourfold}(a), all risk attitudes in the right subplot correspond to $\text{RA}>0$ and therefore risk averse, while those on the left are only risk averse for a certain price range. That is, the four fold pattern is violated in the left subplot. The same trend is exhibited in all four quadrants. This is because, the fourfold pattern is due to the interplay between $\pi(\cdot)$ and $V(\cdot)$ and is valid only when the magnitude of $\pi(\cdot)$ is sufficiently large relative to that of $V(\cdot)$, such that probability distortion dominates \cite{harbaugh2009fourfold}. This corresponds to the right subplots\footnote{The subplot on the right in each quadrant corresponds to the case where individuals are risk neutral in the gain or loss regimes separately, and loss neutral, then $\pi(\cdot)$ alone is sufficient to generate the fourfold pattern.} as well as the left subplots within certain price ranges. 

The implication that we obtain from the analysis of the fourfold pattern of risk attitudes is that the resulting four categories can suitably inform the dynamic pricing strategy in the SMoDS, through the left subplots. That is, it allows a quantification of two qualitative statements (1) the presence of risk seeking passengers gives flexibility in increasing tariffs, and (2) the presence of risk averse passengers requires additional constraints on tariffs.  

\subsection{Strong Risk Aversion over Mixed Prospects}
\label{mixed}

The other implication of the CPT framework is strong risk aversion over mixed prospects. A mixed prospect is defined as an uncertain prospect whose portfolio of possible outcomes involves both gains and losses \cite{kahneman2013prospect, abdellaoui2008tractable}. Clearly, the uncertain prospect is always mixed when $R$ corresponds to its expectation. The strong risk aversion of mixed prospects stems from loss aversion, as the impact of the loss component often dominates its gain counterpart. This implication will be illustrated below in the SMoDS context using two different interpretations. 

The first interpretation follows from Property \ref{existence_lambda}, which essentially states that when $R = \bar{U}$, the subjective utility is strictly negative for a sufficiently large $\lambda$. Therefore, with $R = \bar{U}$ and such a $\lambda$, the uncertain prospect is subjectively perceived as a strict loss. This has been verified numerically using the distributions stated in Section. \ref{distribution} with ${\lambda}^* = 2.25$ as chosen in Table 1. Since the objective utility relative to the expectation is neutral, hence strong aversion is exhibited. 

The second interpretation follows from Property \ref{prob_a_for_mixed}, which essentially states that when Property \ref{existence_lambda} holds, within the tariff range $[\underline{{\lambda}}, \overline{{\lambda}})$, the uncertain prospect is less likely to be accepted by the CPT inclined passengers compared with the rational ones, as $p^s_{\bar{U}} < p^o$.

\begin{figure}[h!]
	\centering
	\includegraphics[width=0.9\textwidth]{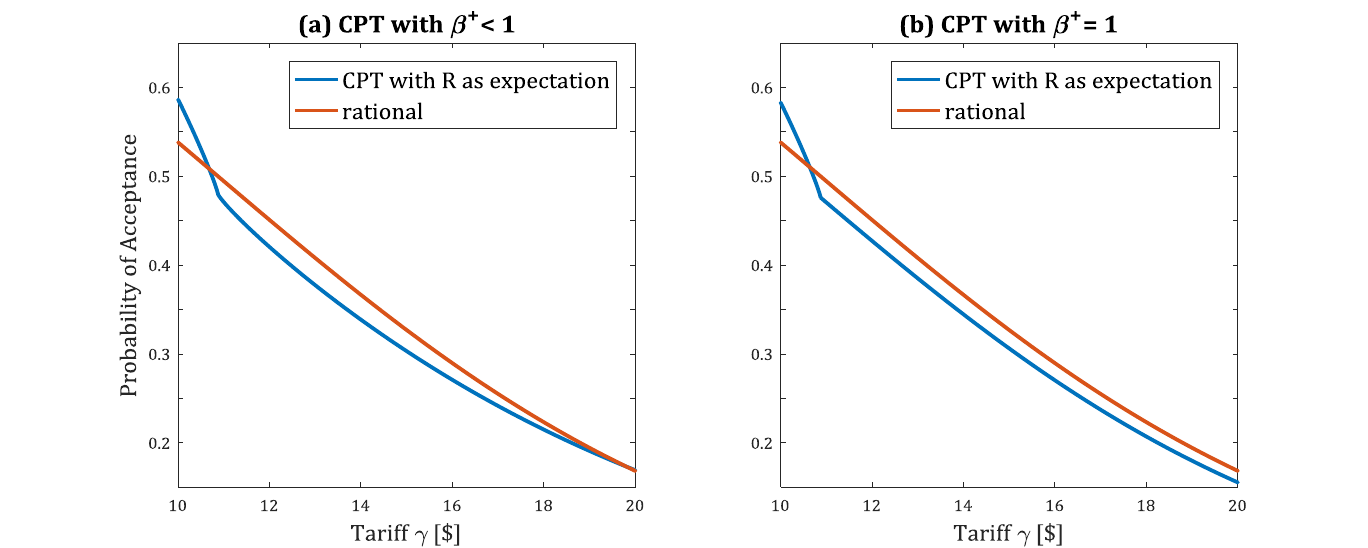}
	\caption{Comparison of $p^s_{\bar{U}}$ and $p^o$. For fair comparison, the tariff range of $\gamma \geq \frac{A^o - \bar{X}}{b}$ is plotted, where the alternative is non-loss.}
	\label{fig:mixed}
\end{figure}
Fig. \ref{fig:mixed} illustrates Property \ref{prob_a_for_mixed} with $f_X(x)$ obeying a Normal distribution. With the numerical values in Table 1, we can compute $\underline{\gamma} \approx \$11$ and $\overline{\gamma} \approx \$20$. It is clear from the left subplot that within this price range, passengers exhibit strong risk aversion over the SMoDS, as the orange curve is strictly above the blue one. It is interesting to note that when ${\beta}^+ = 1$, which corresponds to the case when passengers are risk neutral in the gain regime, $\overline{\gamma} \rightarrow \infty$ (see Fig. \ref{fig:mixed}(b)).  

The implication regarding strong risk aversion over mixed prospects is as follows: As the SMoDS has significant uncertainty, for passengers who regard the expected service quality as the reference, and when the alternative is relatively a non-loss prospect, strong risk aversion is exhibited. Hence the SMoDS is strictly less attractive to these passengers when compared to rational ones. Therefore, the dynamic tariffs may need to be suitably designed by the SMoDS server so as to compensate for these perceived losses. Rebates and subsidies may be a few typical examples. 

\subsection{Self Reference}
\label{self}

\begin{figure}[h!]
	\centering
	\includegraphics[width=0.9\textwidth]{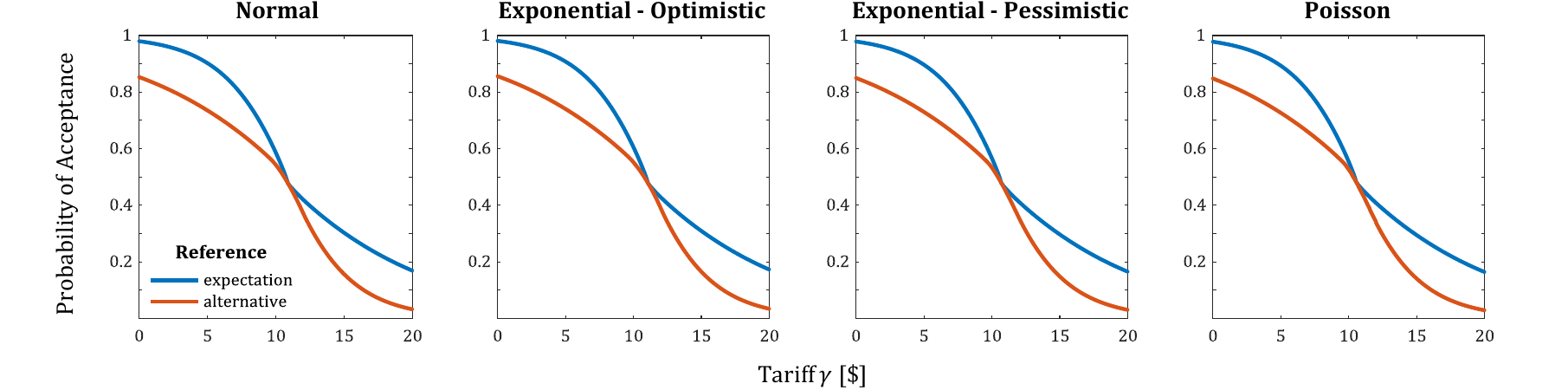}
	\caption{Comparison of $p^s_{\bar{U}}$ with $p^s_{A^o}$ using four different $f_X(x)$ in Section \ref{distribution}. In the truncated Poisson distribution, the parameters are set as ${\lambda}^P = 4$ and $K = 5$.}
	\label{fig:comparison}
\end{figure}

In this section, we compare $p^s_{\bar{U}}$ with $p^s_{A^o}$. The four different distributions defined in (\ref{define_normal})-(\ref{define_Poisson}) are all considered. In each case, how these two probabilities vary with the tariff $\gamma$ were evaluated. The results are shown in Fig. \ref{fig:comparison}.   

Fig. \ref{fig:comparison} illustrates that for all four distributions, $p^s_{\bar{U}} \geq p^s_{A^o}, \forall \gamma$, which implies that the SMoDS is always more attractive when the reference is the expectation of itself, rather than the alternative. $p^s_{\bar{U}} = p^s_{A^o}$ when $\gamma = \frac{A^o - \bar{X}}{b}$ therefore $\bar{U} = A^o$ hence the two reference points coincide. 

The following summarizes the third implication inferred from Fig. \ref{fig:comparison}: $\bar{U}$ is essentially the rational counterpart of the uncertain prospect. Therefore, it could be argued that, when deciding between two prospects, the chance to accept one prospect is always higher if this prospect itself is regarded as the reference, compared with the case where the alternative is considered as the reference. This is due to loss aversion, i.e., $\lambda > 1$, and can be explained thus: When one prospect is regarded as the reference, by definition, it would never be perceived as a loss and therefore not experience the magnified perception out of losses, whereas the alternative may be subject to being regarded as a loss and therefore can experience this skewed perception. In contrast, if the alternative is chosen as the reference, the roles are reversed\footnote{Other effects of CPT due to $\alpha, {\beta}^+, {\beta}^- < 1$ may result in complicated nonlinearities which might alleviate loss aversion. Therefore, this statement is valid when $\lambda$ is sufficiently large, such that loss aversion dominates, which is the case with the CPT parameters listed in Table 1.}. Moreover, the statement is in fact intuitive as those passengers who regard the expectation as the reference have in some sense already subscribed to the SMoDS, hence are naturally inclined to exhibit a higher probability of acceptance and therefore have higher willingness to pay. This partially explains the reason why converting customers from competitors is typically more difficult than maintaining the current customer base. The last observation from Fig. \ref{fig:comparison} is the invariance of the comparison with the underlying probability distributions, which implies that the above implication on self reference are fairly general. 

\subsection{Dynamic Tariff Design}
\label{design}

With the above analytical properties of CPT based passenger behavioral model, we propose the following algorithm for determining the dynamic tariff. As mentioned at the beginning of Section \ref{background}, the goal is for the actual probability of acceptance $p^s_R$ to reach the desired value $p^*$. We note from (\ref{prob_a_binary_objective}) that $p^s_R$ is a function of $U^s_R$ and $A^s_R$, which in turn is a function of $U$ following (\ref{CPT_calculate_discrete}) through (\ref{CPT_calculate}). Finally, (\ref{define_X}) shows that $U$ is a function of $\gamma$. By combining these equations, we can derive the relationship between $p^s_R$ and $\gamma$ as $p^s_R = f(\gamma)$. According to Property \ref{monotonic_static} and \ref{monotonic_dynamic}), $f(\cdot)$ is strictly monotonic. This in turn implies that the desired dynamic tariff that leads to $p^*$ is given by $\gamma = f^{-1}(p^*)$.

\subsection{Other Remarks}
\label{remarks}

In this entire section, the survey data collected from passengers in Singapore \cite{wang2018risk} has been used for the utility coefficients in Table 1. The generality of the above observations and implications can be quantified using the parameter Value of Time (VOT), which equates to $\frac{a_2}{b}$. In the Singapore survey, $\text{VOT} = 0.22$ and $0.77$ [\$/min] for the SMoDS and UberX respectively, and VOT $= 0.40$ [\$/min] for business travelers in the US \cite{Rogoff2014VOT}, both of which are of the same order of magnitude. This implies that the construction of our synthetic data that combines two different sources, one from Singapore and one from the US, is a reasonable excise. 

Another point worth noting is that we have examined the CPT based passenger behavioral model depends on relative pricing rather than the absolute values. Such an examination helps in applying the CPT framework we have proposed and the corresponding observations and implications obtained in this paper in a broader set of problems in the SMoDS. 

\section{Concluding Remarks}
\label{conclusions}

In this paper, we have proposed a dynamic pricing strategy for a SMoDS using Cumulative Prospect Theory, and builds on our previous work in \cite{guan2019dynamicrouting} and \cite{annaswamy2018transactive}. The proposed dynamic pricing strategy together with dynamic routing via the AltMin algorithm \cite{guan2019dynamicrouting}, provide a complete solution to shared mobility on demand that corresponds to an ideal combination of flexibility, convenience, and affordability. The basic principles of CPT and the derivation of the passenger behavioral model in the SMoDS context were described in detail. The three implications of CPT, including the fourfold pattern of risk attitudes, strong risk aversion over mixed prospects, and self reference, on the dynamic pricing strategy of the SMoDS were delineated via computational experiments. The observations and implications obtained in this paper provide a quantitative framework to analyze subjective decision making of passengers in the SMoDS context and can be generalized to a broader set of socio-technical systems.  

Future works will concentrate on the development of $H(\cdot, \cdot | \cdot)$ which is able to achieve robust regulations of $\text{EWT}(t)$ round ${\text{EWT}}^*$, and the investigation of suitable ${\text{EWT}}^*$ scaling that result in an optimal combination of revenue and ridership for the SMoDS platform. The integration of dynamic pricing directly into dynamic routing, and the extension to the case where the server has little information regarding $f_X(x)$ are topics for future investigation as well.




\section*{Acknowledgments}

The authors are grateful to Prof. Jinhua Zhao and Dr. Shenhao Wang from MIT Urban Mobility Lab for valuable suggestions and discussions. This work was supported by the Ford-MIT Alliance. 

\section*{Appendix: Proofs of Properties}

\begin{proof}[Proof of Property \ref{monotonic_static}]
	We prove by definition. $\forall {\gamma}_1, {\gamma}_2 \in {\mathbb{R}}$, ${\gamma}_1 < {\gamma}_2$, and $R \in \mathbb{R}$, we firstly compare $U^s_R({\gamma}_1)$ with $U^s_R({\gamma}_2)$. $\forall u(\gamma)$ such that $u({\gamma}_1) = x+b{\gamma}_1 < R$ or $u({\gamma}_2) = x+b{\gamma}_2 \geq R$, the contribution to $U^s_R(\gamma)$ strictly decrease since $V[u(\gamma)]$ strictly decreases and the weighting remains the same. $\forall u(\gamma)$ such that $u({\gamma}_1) = x+b{\gamma}_1 \geq R$ and $u({\gamma}_2) = x+b{\gamma}_2 < R$, the contribution to $U^s_R(\gamma)$ strictly decreases since $V[u(\gamma)]$ turns from nonnegative to negative and the weighting is positive. Hence $U^s_R({\gamma}_1) > U^s_R({\gamma}_2)$, $A^s_R({\gamma}_1) = A^s_R({\gamma}_2)$, therefore $p^s_R({\gamma}_1) > p^s_R({\gamma}_2)$. Since ${\gamma}_1$ and ${\gamma}_2$ are arbitrarily chosen, $p^s_R(\gamma)$ strictly decreases with $\gamma$.
\end{proof}

\begin{proof}[Proof of Property \ref{monotonic_dynamic}]
	We prove by definition. $\forall {\gamma}_1, {\gamma}_2 \in {\mathbb{R}}, {\gamma}_1 < {\gamma}_2$, and $R = \tilde{x}+b\gamma, \tilde{x} \in \mathbb{R}$, according to (\ref{V_function}), $A^s_R({\gamma}_1) < A^s_R({\gamma}_2)$. To calculate $U^s_R(\gamma)$, all possible outcomes $u(\gamma) = x+b\gamma$ shift the same amount $b|{\gamma}_1-{\gamma}_2|$ as $R $ does, hence the contributions to $U^s_R(\gamma)$ remain the same as both the weighing and $V[u(\gamma)]$ remain the same, therefore $U^s_R({\gamma}_1) = U^s_R({\gamma}_2)$, hence $p^s_R({\gamma}_1) > p^s_R({\gamma}_2)$. Since ${\gamma}_1$ and ${\gamma}_2$ are arbitrarily chosen, $p^s_R(\gamma)$ strictly deceases with $\gamma$. 
\end{proof}

\begin{proof}[Proof of Property \ref{existence_lambda}]
	We prove the case where $U$ is discrete. According to (\ref{CPT_calculate_discrete})-(\ref{V_function}), $U^s_{\bar{U}}(\lambda) = -\Big[\sum_{i=1}^k w_i {(\bar{U}-u_i)}^{{\beta}^-}\Big] \lambda + \sum_{i=k+1}^n w_i {(u_i - \bar{U})}^{{\beta}^+}$. Since $U$ is uncertain, $ -\Big\{\sum_{i=1}^k w_i {(\bar{U}-u_i)}^{{\beta}^-}\Big\} < 0$, one could simply choose ${\lambda}^* = \frac{\sum_{i=k+1}^n w_i {(u_i - \bar{U})}^{{\beta}^+}}{\sum_{i=1}^k w_i {(\bar{U}-u_i)}^{{\beta}^-}} $. The proof of the continuous case follows the same procedure.
\end{proof}

\begin{proof}[Proof of Property \ref{prob_a_for_mixed}]
	Denote $\overline{{\Delta}^o} = A^o - [\bar{X} + b \overline{\gamma}]$ for ease of notation. We firstly prove that there exists a unique $\overline{\gamma}$, such that $\overline{\gamma} > \underline{\gamma}$ and ${\overline{{\Delta}^o}} ^ {{\beta}^+} - U^s_{\bar{U}}  = \overline{{\Delta}^o}$. Moreover, $\forall \gamma \in [\underline{\gamma}, \overline{\gamma})$, $\overline{{\Delta}^o} ^ {{\beta}^+} - U^s_{\bar{U}} > \overline{{\Delta}^o}$. Since $U^s_{\bar{U}} < 0$, and $\overline{\gamma} > \underline{\gamma}$, therefore  $\overline{{\Delta}^o} > 1$. Within the range $\overline{{\Delta}^o} \in (1, \infty)$, $\Big(\overline{{\Delta}^o} - { \overline{{\Delta}^o}} ^ {{\beta}^+}\Big)$ strictly increases, hence there exists a unique $\overline{{\Delta}^o}$, therefore a unique $\overline{\gamma}$, such that ${\overline{{\Delta}^o}} ^ {{\beta}^+} - U^s_{\bar{U}} = \overline{{\Delta}^o}$. In addition, $\forall \gamma \in [\underline{\gamma}, \overline{\gamma})$, $\overline{{\Delta}^o} ^ {{\beta}^+} - U^s_{\bar{U}} > \overline{{\Delta}^o}$ since $\overline{{\Delta}^o} - { \overline{{\Delta}^o}} ^ {{\beta}^+}$ strictly increases. Secondly, since $p^s_{\bar{U}} = \frac{e ^ {U^s_{\bar{U}}}}{e^{U^s_{\bar{U}}} + e^{A^s_{\bar{U}}}} = \frac{1}{1 + e^{A^s_{\bar{U}} - U^s_{\bar{U}}}}$, and $p^o = \frac{e ^ {U^o}}{e^{U^o} + e^{A^o}} = \frac{1}{1 + e^{A^o - U^o}}$, therefore $p^s_{\bar{U}} < p^o \iff (A^s_{\bar{U}} - U^s_{\bar{U}}) > A^o - U^o$. Since $\gamma \geq \underline{\gamma}$, and $R={\bar{U}}$, therefore $A^s_{\bar{U}} = {[A^o - {\bar{U}}]}^{{\beta}^+}$. Since $U^s_{\bar{U}} < 0$, and $U^o = {\bar{U}}$ by definition, hence $(A^s_{\bar{U}} - U^s_{\bar{U}}) > A^o - U^o \iff {[A^o - {\bar{U}}]}^{{\beta}^+} - U^s_{\bar{U}} > A^o - {\bar{U}}$. Since $\gamma \in [\underline{\gamma}, \overline{\gamma})$, the inequality holds.
\end{proof}



\bibliographystyle{plain}

\bibliography{references}


\end{document}